\documentclass{prep-ws-mpla-style}
\usepackage[super]{cite}
\usepackage{graphicx}
\usepackage{amsmath,bm}
\pdfoutput=1	% needed to include PDF graphics
\usepackage{color}
\usepackage{hyperref}

\renewcommand{\vec}[1]{\boldsymbol{#1}}
\newcommand{\diff}{\mathrm{d}}

\newcommand{\eg}{\emph{e.g.}}
\newcommand{\ie}{\emph{i.e.}}
\newcommand{\cf}{\emph{cf.}}

\begin{document}

\markboth{Jan Heisig}{Cosmic-ray antiprotons in the AMS-02 era:~A sensitive probe of dark matter}

%%%%%%%%%%%%%%%%%%%%% Publisher's Area please ignore %%%%%%%%%%%%%%
\catchline{}{}{}{}{}
%%%%%%%%%%%%%%%%%%%%%%%%%%%%%%%%%%%%%%%%%%%%%%%%%%%%%%%%%%%%%%%%%%%

\title{Cosmic-ray antiprotons in the AMS-02 era:\\ A sensitive probe of dark matter}

\author{\footnotesize Jan Heisig}

\address{Centre for Cosmology, Particle Physics and Phenomenology (CP3),\\ Universit\'e catholique de Louvain, Chemin du Cyclotron 2, B-1348 Louvain-la-Neuve, Belgium \\ jan.heisig@uclouvain.be}

\maketitle

\begin{abstract}
Cosmic-ray antiprotons are a powerful tool for astroparticle physics.
While the bulk of measured antiprotons is consistent with a secondary origin,
the precise data of the AMS-02 experiment provides us with encouraging prospects to search for 
a subdominant primary component, \eg~from dark matter. In this brief review, we discuss
recent limits on heavy dark matter as well as a tentative signal from annihilation of dark matter with 
a mass $\lesssim100$\,GeV. We emphasize the special role of systematic errors that can affect the signal. 
In particular, we discuss recent progress in the modeling of secondary production cross sections and 
correlated errors in the AMS-02 data, the dominant ones originating from uncertainties in the cross sections 
for cosmic-ray absorption in the detector.
\end{abstract}

\ccode{PACS Nos.: 96.50.S, 95.35.+d}

%===================================================================
\section{Introduction}\label{sec:intro}
%===================================================================

Since their first observations over 100 years ago~\cite{Hess:1912srp} much has been learned in the study of
cosmic rays. Modern experiments (both direct and air-shower observations) provide invaluable information about the composition and energy spectra 
of cosmic rays over a huge energy range shedding light on their origin and propagation environment~\cite{Strong:2007nh,Gaisser:2013bla,Amato:2017dbs}.
The AMS-02 experiment on-board the International Space Station has measured the fluxes of a large set of cosmic-ray nuclei in the GeV to TeV
range~\cite{Aguilar:2015ooa,Aguilar:2015ctt,Aguilar:2016kjl,Aguilar:2016vqr,Aguilar:2017hno,Aguilar:2018njt}
that are of Galactic origin.
The most abundant species are \emph{primary} cosmic rays, such as protons ($\sim90\%$), He ($\sim10\%$) and C, O ($\sim1\%$),
that are accelerated to high energies through diffusive shock acceleration, \eg~in the environment of supernova remnants.
In contrast, the origin of `fragile' nuclei, such as Li, Be, B that are easily destroyed in stellar processes, are of \emph{secondary} nature.
The same applies to antiprotons, as any significant amount of anti-baryons is absent in stellar matter.
Secondaries are produced in collisions of 
primaries with the interstellar medium, \ie~spallation or antiparticle production processes.

Dark matter constitutes an additional primary source of antiprotons in the Galaxy, provided its interaction supports pair-annihilation
into standard-model particles (see \eg~Ref.~\refcite{Bertone:2004pz} for a review). 
The blueprint of such a dark-matter candidate is a weakly interacting massive particle, with theoretically appealing features~\cite{Jungman:1995df}. In fact,
antiprotons have been suggested as a possible probe of such a candidate over 30 years ago~\cite{Silk:1984zy,Stecker:1985jc}.
Since then they have become a standard tool for indirect detection of dark matter~\cite{Bergstrom:1999jc,Donato:2003xg,Bringmann:2006im,Donato:2008jk,Fornengo:2013xda,Hooper:2014ysa,Pettorino:2014sua,Boudaud:2014qra,Cembranos:2014wza,Cirelli:2014lwa,Bringmann:2014lpa,Giesen:2015ufa,Jin:2015sqa,Evoli:2015vaa,Cuoco:2016eej,Cuoco:2017iax,Reinert:2017aga,Cui:2018klo,Cuoco:2019kuu,Heisig:2020nse}.

However, the interpretation of cosmic-ray data is complicated by several aspects that are interconnected.
First, the propagation of cosmic-rays through the Galaxy is a highly non-trivial process whose modeling is still lacking a sufficiently predictive theoretical framework,
rending a fit to the data the only way to determine its parameters. Secondly, the modeling of primary sources faces a similar problem. While diffusive shock acceleration generically 
predicts a simple power-law behavior for the injection spectra, observations~\cite{Ahn:2010gv,Adriani:2011cu,Aguilar:2015ooa,Aguilar:2015ctt} necessitate 
the use of different spectral indices (for protons and heavier nuclei) 
as well as (possibly)~\cite{Serpico:2018lkb} the introduction of 
spectral breaks. This has led to an increase in the number of model parameters that, again, can only be constrained by a fit to the data.
Thirdly, several sources of systematic uncertainties are at play that often arise from poorly constrained nuclear processes (in its relevant kinematic regime for the cosmic-ray analysis). 
Notably, these uncertainties have been shown to exhibit sizeable correlations calling out for a careful assessment~\cite{Reinert:2017aga,Cuoco:2019kuu,Derome:2019jfs,Boudaud:2019efq,Heisig:2020nse}.

With the  AMS-02 experiment, cosmic-ray physics has entered a precision era. For the first time, it provides measurements 
at a percent-level precision that allows for a welcomed redundancy in the constraints of currently considered propagation models.
Strong efforts have been made to exploit such a wealth of data
to draw solid conclusions on the existence of a primary antiproton contribution from dark matter. However, tackling the inverse problem of cosmic-ray propagation remains a
major challenge. It requires the disentangling of effects from the modeling of astrophysical cosmic-ray sources, cosmic-ray propagation, involved uncertainties and a possible dark matter signal. 
While this effort is subject to ongoing research, this article reviews recent results on the derivation of robust limits on heavy dark matter~\cite{Cuoco:2017iax} as well as a potential hint for
dark matter with a mass $\lesssim 100\,$GeV~\cite{Cuoco:2016eej,Cui:2016ppb,Cuoco:2017rxb,Reinert:2017aga,Cui:2018klo,Cuoco:2019kuu,Cholis:2019ejx,Lin:2019ljc,Heisig:2020nse}. We discuss possible degeneracies between propagation and dark-matter parameters and various sources of (correlated) systematic uncertainties.

The remainder of this paper is organized as follows. In section~\ref{sec:crde} we review cosmic-ray propagation and the ways to constrain its parameters. In sections~\ref{sec:limits} and \ref{sec:exc} we discuss  
limits on dark matter and a tentative dark-matter excess, respectively. We take a closer look into systematic uncertainties in section~\ref{sec:uncer} before summarizing in section~\ref{sec:sum}.

%===================================================================
\section{Cosmic-ray propagation}\label{sec:crde}
%===================================================================

%-----------------------------------------------------------------------------------------------------------
\subsection{Diffusion equation}
%-----------------------------------------------------------------------------------------------------------

The propagation of charged cosmic rays is characterized by numerous deflections on the turbulent magnetic fields in our Galaxy 
(see \eg~Refs.~\refcite{Strong:2007nh,Amato:2017dbs} and references therein).
They render propagation to be a diffusive process 
with a typical residence time
of several million years in the diffusion volume.
A state-of-the-art description is provided by the diffusion equation for the particle density $\psi_i$ of species $i$ per volume and absolute 
value of momentum $p$:~\cite{Ginzburg:1990sk,Strong:2007nh}
\begin{equation}
\begin{split}
  \label{eq:PropagationEquation}
  \frac{\partial \psi_i (\vec{x}, p, t)}{\partial t} = \;&
    q_i(\vec{x}, p) +  
    \vec{\nabla} \cdot \left(  D_{xx} \vec{\nabla} \psi_i - \vec{V} \psi_i \right)
     +  \frac{\partial}{\partial p} p^2 D_{pp} \frac{\partial}{\partial p} \frac{1}{p^2} \psi_i   \\
   &- \frac{\partial}{\partial p} \left( \frac{\diff p}{\diff t} \psi_i  - \frac{p}{3} (\vec{\nabla \cdot V}) \psi_i \right) 
    - \frac{1}{\tau_{f,i}} \psi_i - \frac{1}{\tau_{r,i}} \psi_i\,.
    \end{split}
\end{equation} 
Here, $q_i(\bm{x}, p)$ is the source term (for primary and secondary sources). Terms 
proportional to $D_{xx},  \vec{V}$, and $D_{pp}$ correspond to spatial diffusion, 
convection and reacceleration, respectively. The second line of 
eq.~\eqref{eq:PropagationEquation} represents the momentum gain or 
loss rate~$\propto {\diff p}/{\diff t}$, adiabatic energy losses 
$\propto \bm{\nabla \cdot V}$, and the catastrophic loss of particles by fragmentation and radioactive 
decay, $\propto 1/\tau_{f,i}$ and $1/\tau_{r,i}$, respectively. We discuss the various terms in more detail below.

Equation~\eqref{eq:PropagationEquation} constitutes a coupled set of partial differential equations.
As fragmentation and decay of heavier nuclei provide a source of lighter ones, the chain of coupled
equations is usually solved starting from heavier to lighter nuclei (and from primaries to secondaries).
In practice, one often employs the steady-state solution and adopts the free escape boundary condition,
\ie~the condition of vanishing densities at the boundaries of the propagation volume. The latter is typically chosen
to be a cylindrical volume with a radius $r\sim20\,$kpc and half-height $z_\text{h}\sim2\!-\!10\,$kpc~\cite{Evoli:2019iih,Weinrich:2020ftb} centered around
our Galaxy.
The equation can be solved fully numerically, using computer codes such as \textsc{Galprop}~\cite{Strong:1998fr,Strong:2015zva}, \textsc{Dragon}~\cite{Evoli:2008dv,Evoli:2017vim} and \textsc{Picard}~\cite{Kissmann:2014sia} or by (semi-) analytical methods, as implemented \eg~in~\textsc{Usine}~\cite{Putze:2010zn,Maurin:2018rmm}.

\smallskip
\textbf{Primary source terms.} Diffusive shock acceleration implies a power-law behavior of primary injection spectra (see \eg~Ref.~\refcite{Drury:1983zz}).
However, AMS-02 data~\cite{Aguilar:2015ooa,Aguilar:2015ctt} (in parts confirming earlier indications~\cite{Ahn:2010gv,Adriani:2011cu}) suggest the introduction of spectral breaks.
This concerns a high-rigidity (${\cal R}\sim300$\,GV) and low-rigidity (${\cal R}\lesssim 10$\,GV) break~\cite{Trotta:2010mx,Evoli:2015vaa,Johannesson:2016rlh}, often modeled with smoothened transitions~\cite{Korsmeier:2016kha}. (Here, ${\cal R} = p/Z$, with $Z$ being the particle's charge number.)
The spatial distribution of primary sources is highly concentrated around the Galactic disk, $\propto \exp(-|z|/z_0)$, with a characteristic half-hight of $z_0\simeq0.2\,$kpc.

\smallskip
\textbf{Spatial diffusion.} The spatial diffusion coefficient $D_{xx}$ constitutes the central piece of the diffusion equation~\cite{Ginzburg:1964,Ginzburg:1990sk}.
A commonly made assumption is a homogeneous and isotropic coefficient.
In its simplest form, it is parametrized by a single power-law in rigidity:
\begin{equation}
\label{eq:diff}
D_{xx} = \beta^\eta D_0 {\cal R}^\delta
\end{equation} 
with $\beta$ being the particle's velocity (in units of $c=1$) and $\eta=1$.
However, with the precision of AMS-02 data, for the first time, one can distinguish whether the breaks seen in the primary fluxes arise from the injection spectra or propagation, see \eg~Ref.~\refcite{Serpico:2018lkb}. In fact, AMS-02 data on Li,\,Be,\,B~\cite{Aguilar:2018njt} suggest to assign the high-rigidity break (${\cal R} \sim 300\,$GV) to the diffusion coefficient rather than the injection spectra~\cite{Genolini:2017dfb}. 
Several microphysical mechanisms have been proposed that support such a behavior~\cite{Blasi:2012yr,Aloisio:2015rsa}.
Furthermore, the inclusion of the low-rigidity break (${\cal R} \lesssim 10\,$GV) in diffusion and/or $\eta\neq1$ has recently shown to provide good fits to AMS-02 data on B/C~\cite{Genolini:2019ewc} as well as other secondary-to-primary ratios~\cite{Weinrich:2020cmw}. Such a break may arise as a consequence of damping of small-scale magnetic turbulences (see \eg~Ref.~\refcite{Blasi:2012yr}).

\smallskip
\textbf{Convection and reacceleration.} Convective winds generated by astrophysical sources are directed perpendicular to the Galactic plane. They are parametrized by the convection velocity, $v_\text{c}$. 
Reacceleration in the turbulent magnetic fields introduces diffusion in momentum space. It is linked to the spatial diffusion coefficient via the velocity of Alfven magnetic waves, $v_\text{A}$.~\cite{Ginzburg:1990sk,1994ApJ...431..705S} Both processes are most relevant at low rigidities. 

\smallskip
\textbf{Secondary source terms.}  The secondary source terms depend on the 
fluxes $\phi_i$ of projectile nuclei $i$ and the number density $n_{\mathrm{ISM},j}$
of target nuclei $j$ in the interstellar medium. To obtain the secondary antiproton source
spectrum the energy-differential cross section, $\diff \sigma_{i,j}/\diff T_{\bar{p}} $, has to be convoluted
with the energy-dependent flux $\phi_i(T_i)$:~\cite{Shen:1968zza,Bottino:1998tw}
\begin{eqnarray}
	\label{eq:sec_sourceTerm}
	q_{\bar p}^{ij}({\bm x},T_{\bar{p}}) &=& \int\limits_{T_{\rm th}}^\infty \diff T_i \,\, 
    	4\pi \,n_{\mathrm{ISM},j}({\bm x}) \, \phi_i  (T_i) \, \frac{\diff\sigma_{ij}}{\diff T_{\bar{p}}}(T_i , T_{\bar{p}})\,,
\end{eqnarray}
where $T$ denotes the kinetic energy and the threshold for antiproton production is $T_\text{th} = 6 m_p$ (assuming no anti-matter in the final state). 
The dominant contribution for antiproton production comes from proton-proton, proton-He and He-proton. See section~\ref{sec:xs} for more details.

For the spallation of heavier into lighter nuclei (responsible for the production of Li, Be, B) no convolution has to be performed as, to first approximation~\cite{1995ApJ...451..275T},
the kinetic energy per nucleon is constant, $T_i/A_i = T_k/A_k$, where $k$ is the secondary nucleus and $A$ the nucleon number.

\smallskip
\textbf{Dark-matter source term.} If dark matter is made of a self-annihilating particle it introduces a 
primary source term for antiprotons:
\begin{equation}
  \label{eq:DM_source_term}
  q_{\bar{p}}^{(\mathrm{DM})}(\bm{x}, E_\mathrm{kin}) = 
  \frac{1}{2} \left( \frac{\rho(\bm{x})}{m_\mathrm{DM}}\right)^2  
  \sum_f \left\langle \sigma v \right\rangle_f \frac{\diff N^f_{\bar{p}}}{\diff E_\mathrm{kin}} ,
\end{equation}
where $m_\mathrm{DM}$ and $\rho(\bm{x})$ denotes the dark-matter mass and density 
profile in the Galaxy, respectively, $\left\langle \sigma v \right\rangle_f$ the velocity averaged annihilation 
cross section and $\diff N^f_{\bar{p}}/\diff E_\mathrm{kin}$ the corresponding 
antiproton energy spectrum per dark-matter annihilation.\footnote{Note that the factor $1/2$ in eq.~\eqref{eq:DM_source_term}
corresponds to the case of a self-conjugate dark-matter candidate, \eg~a Majorana fermion. The corresponding factor is $1/4$ otherwise.}
For annihilation into all pairs of standard-model particles, annihilation spectra have been computed and made publicly available for a large range
of dark-matter masses~\cite{Cirelli:2010xx}. They can also be computed for arbitrary (and mixed) final states using the automated numerical tool \textsc{MadDM}~\cite{Ambrogi:2018jqj}
(utilizing \textsc{Pythia}8~\cite{Sjostrand:2014zea} for showing and hadronization).
For a given density profile $\rho$ and annihilation channel $f$ the source introduces the two dark-matter model parameters, $m_\mathrm{DM}$ and $\left\langle \sigma v \right\rangle_f$.

While the local dark-matter density at the solar position is constrained at a level of $\sim30\%$~\cite{Salucci:2010qr}, 
the density profile towards the Galactic center is only loosely constrained by data allowing for both
`cuspy'  profiles, such as Navarro-Frenk-White~\cite{Navarro:1995iw} or Einasto~\cite{Einasto:1965czb} as well as `cored' profiles such as Burkert~\cite{Burkert:1995yz}.
We briefly discuss the role of different choices for cosmic-ray antiproton observation in section~\ref{sec:limits}.

Finally, we would like to mention that other exotic astrophysical sources of antiprotons have been considered in the literature, see \eg~Ref.~\refcite{Blasi:2009bd,Kohri:2015mga}.

%-----------------------------------------------------------------------------------------------------------
\subsection{Solar modulation}
%-----------------------------------------------------------------------------------------------------------

For observations near Earth, the inverse problem of cosmic-ray propagation is further complicated by a
variety of transport processes in the heliosphere, collectively referred to as \emph{solar modulation} (see \eg~Ref.~\refcite{Potgieter:2013pdj} for a review).
Upon entering the solar system cosmic rays interact with the turbulent solar wind and heliospheric magnetic field.
This causes a suppression of the measured flux near Earth compared to the interstellar one, in particular at small rigidities. For (anti)protons
the effect becomes important for rigidities below a few tenths of GV. The strength of solar modulation is correlated with the solar activity 
which undergoes an 11-year cycle. Several numerical codes have been developed to
solve the transport equation for heliospheric models on different levels of sophistication~\cite{Kappl:2015hxv,Vittino:2017fuh,Aslam:2018kpi,Boschini:2019ubh,Kuhlen:2019hqb}.
Recent progress in constraining solar-modulation models has been due to the direct
measurement~\cite{Stone_VOYAGER_CR_LIS_FLUX_2013} of interstellar fluxes by the Voyager I spacecraft that has left the heliosphere
in 2012 as well as the time-dependent
fluxes released by PAMELA~\cite{Adriani:2013as,Adriani:2016uhu}  and AMS-02~\cite{Aguilar:2018wmi,Aguilar:2018ons}.

For practical purposes, in cosmic-ray studies, one often adopts the force-field approximation~\cite{Gleeson:1968zza,Fisk_SolarModulation_1976},
describing solar modulation by a single parameter: the solar-modulation potential. Deviations from this
simple picture have been captured \eg~by a charge-, rigidity- and/or time-dependent solar-modulation potential~\cite{Cholis:2015gna,Tomassetti:2017hbe,Gieseler:2017xry,Cholis:2020tpi}.
For (anti)protons that are subject to this review deviations from the force-field approximation 
become apparent at rigidities below roughly 5\,GV (see \eg~Ref.~\refcite{Cuoco:2019kuu}).

%-----------------------------------------------------------------------------------------------------------
\subsection{Constraining propagation}
%-----------------------------------------------------------------------------------------------------------

Constraining the diffusion model requires knowledge over at least
one primary and secondary cosmic-ray species. As the measured primary spectra (corrected for solar modulation; see above)
are, to first approximation, identical to the ones entering the source term for secondaries, we can 
infer the effect of propagation by the comparison of secondaries to primaries. Approximately, $\psi_\text{s}/\psi_\text{p}\propto D_{xx}^{-1} \propto {\cal R}^{-\delta}$, in the case of simple
power-law diffusion.
The standard secondary-to-primary ratio to constrain diffusion is B/C~\cite{Maurin:2001sj,Moskalenko:2001ya,Maurin:2010zp,Genolini:2015cta,Derome:2019jfs,Genolini:2019ewc}, but also Li/C, Be/C, Li/O, Be/O, B/O, N/O, $^3$He/$^4$He as well as $\bar p / p$ have been considered~\cite{Evoli:2008dv,Korsmeier:2016kha,Johannesson:2016rlh,Wu:2018lqu,Boschini:2019gow,Evoli:2019wwu,Weinrich:2020cmw}. A commonly made assumption is the universality of diffusion, requiring the validity of a model from (anti)protons to O. This assumption is, however, subject to ongoing scrutiny (see \eg~discussion in Refs.~\refcite{Johannesson:2016rlh,Weinrich:2020cmw}).

Note that the diffusion parameter inference from secondary-to-primary flux ratios introduces a high level of degeneracy between the normalization of the diffusion coefficient and the halo size, $z_\text{h}$, consequently leading to  rather weak constraints on either of these quantities individually. This is relevant for dark-matter searches -- their sensitivity is strongly affected by $z_\text{h}$. We will briefly comment on this aspect in section~\ref{sec:limits}. Independent constraints on $z_\text{h}$ can be derived from radioactive secondaries like $^{10}$Be, see \eg~Refs.~\refcite{Evoli:2019iih,Weinrich:2020ftb} for a recent account.

%===================================================================
\section{Constraints on dark matter}\label{sec:limits}
%===================================================================

Using cosmic-ray antiprotons for probing dark-matter annihilation in our Galaxy has a long history. 
In particular, the increasing level of precision of data from balloon- and space-borne experiments, like 
BESS~\cite{Orito:1999re,Maeno:2000qx}, AMS~\cite{Aguilar:2002ad}, BESS-Polar~\cite{Abe:2011nx}, PAMELA~\cite{Adriani:2010rc,Adriani:2012paa} and, eventually, AMS-02~\cite{Aguilar:2016kjl} has established this channel as an important tool to place constraints on dark-matter models~\cite{Bergstrom:1999jc,Donato:2003xg,Bringmann:2006im,Donato:2008jk,Fornengo:2013xda,Hooper:2014ysa,Pettorino:2014sua,Boudaud:2014qra,Cembranos:2014wza,Cirelli:2014lwa,Bringmann:2014lpa,Giesen:2015ufa,Jin:2015sqa,Evoli:2015vaa,Cuoco:2016eej,Cuoco:2017iax,Reinert:2017aga,Cui:2018klo}.

A central challenge in the interpretation of the data has been the diffusion model uncertainties. 
In the pre-AMS-02 era, employing the MIN/MED/MAX benchmark scenarios~\cite{Donato:2003xg} 
(consistent with B/C data at that time~\cite{Maurin:2001sj}), these led to uncertainties in the upper limits on annihilation cross sections
reaching up to three orders of magnitude~\cite{Fornengo:2013xda}. 

With the AMS-02 antiproton data, uncertainties in the measured fluxes are reduced to a few percent over a wide range of rigidities~\cite{Aguilar:2016kjl}
providing sensitivity to a primary dark-matter contribution to antiprotons as low as around 10\%.
However, not only does the new data promote dark-matter searches to a new level of sensitivity, it has also 
brought forward cracks in the standard minimal scenario, requiring a refinement of the propagation model.
This, in turn, comes at the price of increasing the number of free parameters associated with the modeling of propagation (and/or sources).
Accordingly, care has to be taken before assigning a possible anomaly to dark-matter annihilation -- or considered
from the viewpoint of imposing constraints in the absence of an observed excess: We have to make sure that the 
dark-matter hypothesis to be excluded is not only incompatible with data for certain propagation parameters but for all 
choices that are allowed by current data, thereby exploring possible degeneracies in the parameter space. 
This task calls out for a global fit of both propagation and dark-matter parameters where the former
are treated as nuisance parameters.

In Ref.~\refcite{Cuoco:2017iax} we have performed such an analysis. As opposed to constraining the propagation model
by an independent measurement of B/C, in this analysis, we choose a minimal set of primary and secondary fluxes 
that allows us to constrain both the propagation and dark-matter model while not relying on the assumption of universality.
It includes the proton~\cite{Aguilar:2015ooa} and He~\cite{Aguilar:2015ctt} fluxes and the antiproton-to-proton flux ratio ($\bar p/p$)~\cite{Aguilar:2016kjl}. 
We will refer to this setup as the `minimal network' scenario in the following.

The analysis assigns two (smoothened) breaks to the primary injection spectra allowing for individual 
slopes for proton and He and, accordingly, no break in diffusion. We employ convection and reaccelaration
and use \textsc{Galprop} for the numerical solution of the diffusion equation.
Solar modulation is modeled by the force-field approximation constrained by additional proton and He data from Voyager~\cite{Stone_VOYAGER_CR_LIS_FLUX_2013}.
However, for the considered range of dark-matter masses (200\,GeV--50\,TeV) uncertainties from solar modulation have little impact on the results.

\begin{figure}[t]
\vspace{0.5cm}
\centering
\setlength{\unitlength}{1\textwidth}
\begin{picture}(1,0.313)
 \put(0.00,-0.01){\includegraphics[width=1\textwidth]{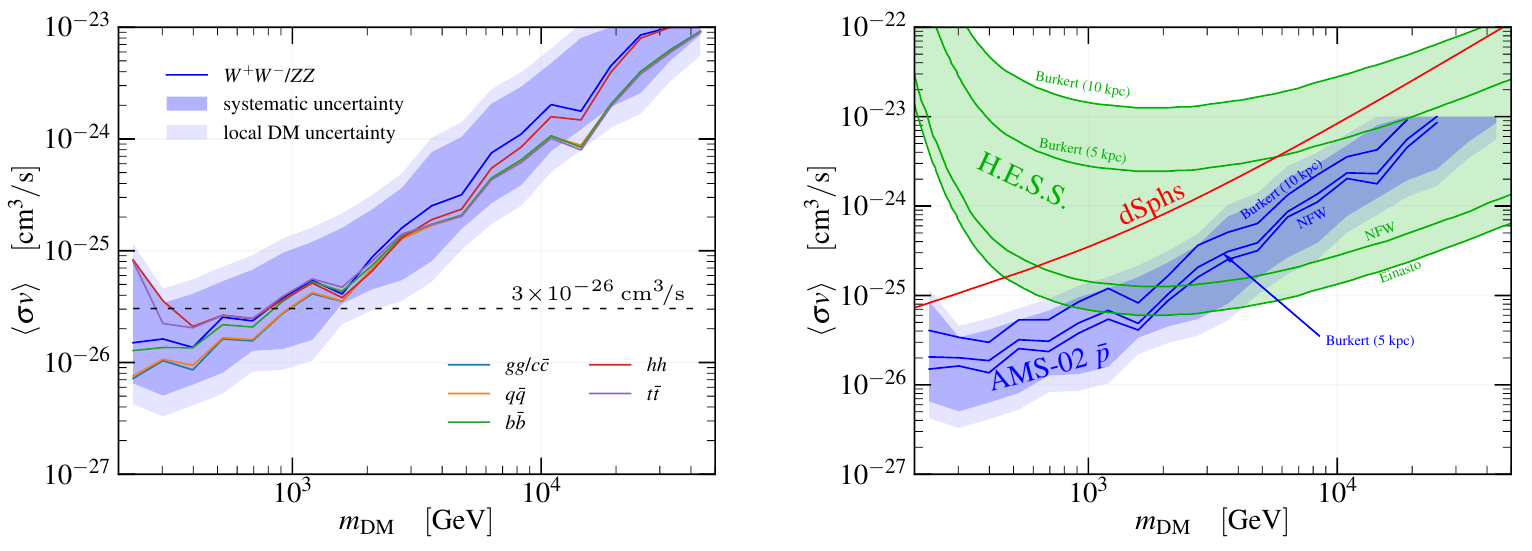}}
\end{picture}
\caption{
Upper limits at 95\% C.L. on the dark-matter annihilation cross section.
Left panel:~Limits for non-leptonic annihilation channels. The dark and light blue shaded bands
provide an estimate for additional systematics and the uncertainty from the local dark-matter density (linearly added) for annihilation into 
$W$-bosons.
Right panel: Comparison of limits from cosmic-ray antiprotons with gamma-ray observations of the Galactic center by H.E.S.S.~\cite{Abdallah:2016ygi} and dwarf spheroidal galaxies by Fermi-LAT~\cite{Fermi-LAT:2016uux} (dSphs) for annihilation into 
$W$-bosons and for various choices of the dark-matter density profile. 
The figure is taken from Ref.~{\protect\refcite{Cuoco:2017iax}}.
}
\label{fig:limits}
\end{figure}

For the derivation of cross-section upper limits we employ a frequentist analysis and construct the profile likelihood as a function of the dark-matter parameters by
profiling over all propagation parameters.
This is a non-trivial task. In particular, it requires careful exploration of the best-fit regions in the propagation parameter space 
for annihilation cross sections around the exclusion limit to be derived. Note that an insufficient parameter sampling in this region would result in a too strong limit.
The results for annihilation into a pair of $W$-boson as well as other non-leptonic channels are shown in 
figure~\ref{fig:limits}. For the $WW$ and $bb$ channel, they exclude the thermal cross section, $\left\langle \sigma v \right\rangle\sim 3\times 10^{-26}\,\text{cm}^3/\text{s}$,
for masses up to 800\,GeV. Note that very similar results have been obtained in Ref.~\refcite{Reinert:2017aga}, in the region 200\,GeV--3\,TeV considered in both analyses.
That study uses the proton flux (instead of $\bar p/p$) and constrains propagation by the independent measurement of the B/C flux ratio.

Figure~\ref{fig:limits} also provides information about further uncertainties that go beyond the uncertainties on the propagation parameters
already taken into account in the course of profiling. These concern the parametrizations of secondary antiproton production cross sections 
as well as changes in the propagation setup, like setting individual key parameters in the fit to fixed (extreme) values. Most significantly, it concerns a fixed half-height, set to $z_\text{h}=2\,$kpc and $z_\text{h}=10\,$kpc.
In almost the entire range the latter two choices provide the upper and lower boundary, respectively, of the dark blue shaded band that represents the envelope of all chosen setups.
However, in the light of recent analyses of the fluxes of unstable secondary cosmic-ray nuclei~\cite{Weinrich:2020ftb,Evoli:2019iih} these values already appear somewhat extreme.

Unlike dark-matter searches in gamma rays that utilize the morphology of the signal to discriminate signal from background, 
cosmic-ray observations are much less sensitive to the dark-matter density profile in the Galaxy. This is quantified in the right panel of figure~\ref{fig:limits}
where the cosmic-ray limits are shown for the default (cuspy) Navarro-Frenk-White profile as well as for a (cored) Burkert profile with a core radius of 5 and 10\,kpc. The difference 
in the limits between the former and the latter amounts to a factor of around 2. 
In comparison, the limits from gamma-ray observation of the Galactic center by H.E.S.S.~\cite{Abdallah:2016ygi} vary by 2 orders of magnitude between these choices (\cf~green curves
and shaded region).

%=====================
%    \                                           |
%      \                                         |
%        \                                       |
%                                      \         |
%                                        \       |
%                                          \     |
%=====================

\smallskip

So far we have parametrized the dark-matter signal by its mass and cross section assuming 
100\% annihilation into single final state channels. But we have not derived implications for realistic 
particle physics models of dark matter. 

The arguably most literal realization of a weakly interacting massive particle is obtained in the framework of
minimal dark matter~\cite{Cirelli:2005uq} which supplements the standard model by just an $SU(2)$ multiplet. This theoretically 
appealing model comes with only one free parameter (the tree-level dark-matter mass) which, in principle, can be fixed by the relic density constraint. 
Notably, the fermion doublet, triplet or quintuplet (with hypercharge 1/2, 0 and 0, respectively)
are of particular interest. The former ones represent limiting cases of a supersymmetric standard model with a pure higgsino and wino dark matter, respectively.
The latter is the simplest representation where dark matter is stable without imposing an additional $Z_2$ symmetry.
For this model, indirect detection constitutes the prime search strategy. The cosmologically preferred range of masses appears mostly out-of-reach for upcoming collider searches, while direct detection 
cross sections vanish at tree level.

%=====================
%    \                                           |
%      \                                         |
%        \                                       |
\begin{figure}[t]
\vspace{0.5cm}
\centering
\setlength{\unitlength}{1.0\textwidth}
\begin{picture}(1,0.322)
 \put(0.00,-0.01){\includegraphics[width=1\textwidth]{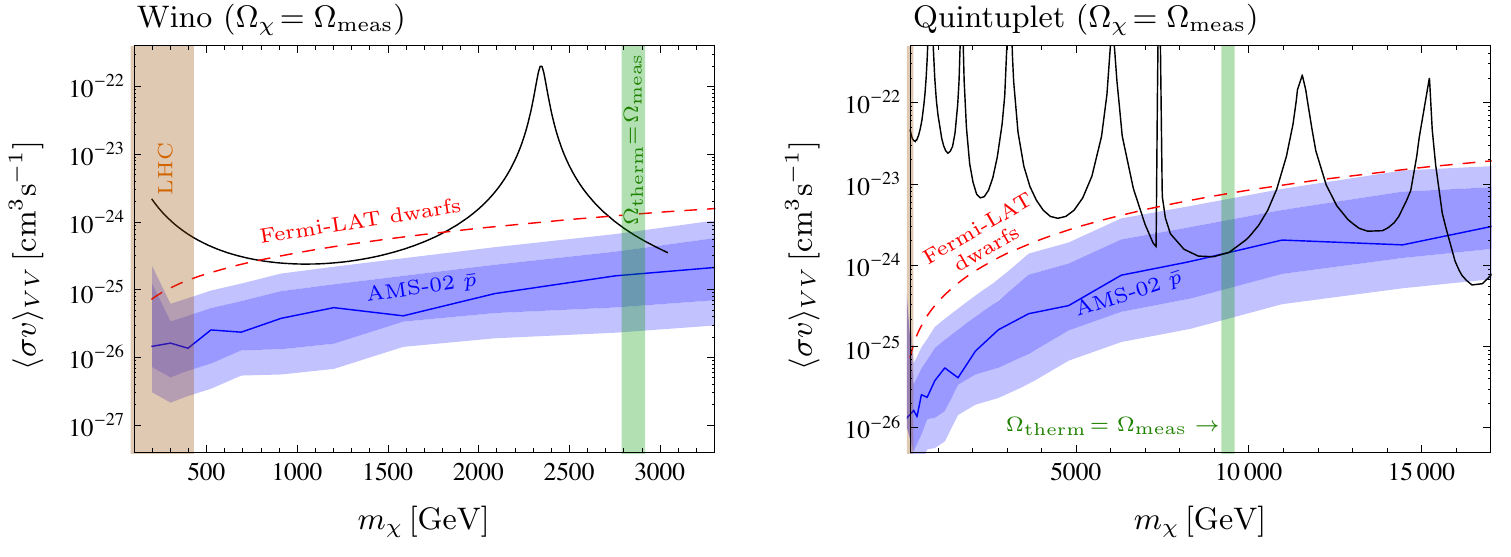}}
\end{picture}
\caption{
95\% CL exclusion limits on the annihilation cross section for minimal dark matter. The left and right panels show the
fermion triplet (wino) and fermion quintuplet, respectively. The blue curves show the 
upper limits from AMS-02 antiprotons. The dark and light blue shaded bands indicate 
the systematic uncertainties and uncertainties from the local dark-matter density (added linearly), respectively. 
The solid black curves show the cross-section prediction within the model.
The vertical green shaded bands (around 2850\,GeV and 9.4\,TeV, respectively) denote the cosmologically preferred regions.
For the limit setting the candidate's relic density is set to the observed value in the entire range. 
The figure is taken from Ref.~{\protect\refcite{Cuoco:2017iax}}.
}
\label{fig:MDM}
\end{figure}
%                                      \         |
%                                        \       |
%                                          \     |
%=====================

In all three cases, annihilation into a pair of electroweak vector boson, $VV=WW, ZZ, Z\gamma $, dominates the dark-matter induced antiproton source term.
Note that its cross section is significantly enhanced due to non-perturbative effects (Sommerfeld enhancement). 
We show the corresponding limits (blue solid curves and shaded bands) and model predictions~\cite{Hryczuk:2011vi,Cirelli:2015bda} (black solid curves) for the triplet and quintuplet in triplet or quintuplet in figure~\ref{fig:MDM}.
In both cases, the cosmologically preferred regions~\cite{Cirelli:2007xd,Cirelli:2015bda} (satisfying the relic density constraint through thermal freeze-out; vertical green shaded bands) are challenged by observations of cosmic-ray antiprotons. In particular, in the case of the wino,
consistency with data requires the full exploitation of the uncertainty band towards the most conservative edge. The higgsino case (not shown here; see Ref.~{\protect\refcite{Cuoco:2017iax} for details)
is less constrained, leaving the cosmologically preferred region unchallenged.

Note that the model is also in tension with results of searches for gamma-line signatures in observations of the Galactic center~\cite{Abramowski:2013ax}. However, this conclusion only holds for the choice of cuspy density profile, \cf~the related discussion above. For cored profiles, these searches are not sensitive to the model~\cite{Cuoco:2017iax}. The situation is, however, expected to change with the upcoming Cherenkov Telescope Array experiment~\cite{Consortium:2010bc} as recently projected in Ref.~\refcite{Rinchiuso:2020skh} (see also~Ref.~\refcite{Hryczuk:2019nql}).

%===================================================================
\section{The antiproton excess}\label{sec:exc}
%===================================================================

While cosmic-ray antiprotons allow us to place strong limits on heavy dark matter, $m_\text{DM}>200\,$GeV,
the constraints significantly weaken towards smaller masses. In fact, data supports a preference for
a dark-matter contribution in this region. This spectral feature was first found in Ref.~\refcite{Cuoco:2016eej} employing the minimal network scenario
(\cf~section~\ref{sec:limits}) and Ref.~\refcite{Cui:2016ppb} using B/C to constrain propagation.
Subsequently, the excess has been analyzed by several groups using various analysis setups~\cite{Cuoco:2017rxb,Reinert:2017aga,Cui:2018klo,Cuoco:2019kuu,Cholis:2019ejx,Lin:2019ljc,Heisig:2020nse}.
While the significance of the excess is subject to controversies, ranging from around $1\sigma$ to above $5\sigma$ in the 
aforementioned studies, a common picture of the preferred dark-matter properties has, in fact, emerged.
It hints at a dark-matter mass of $40\sim\!130\:\text{GeV}$ and an annihilation cross section around the thermal one, $\langle \sigma v\rangle\sim 10^{-26}\:\text{cm}^2\text{s}^{-1}$. 
Following the analysis setup of Ref.~\refcite{Cuoco:2016eej}, in Ref.~\refcite{Cuoco:2017rxb} we have performed joint fits
for a variety of non-leptonic channels.
The left panel of figure~\ref{fig:fitreg} shows the respective best-fit regions in the dark-matter mass versus cross-section plane.
The displayed channels all provide a similar improvement of the fit (formally above $4\sigma$)
except for $t\bar t$ performing somewhat worse (around $3\sigma$).

%=====================
%    \                                           |
%      \                                         |
%        \                                       |
\begin{figure}[t]
\centering
\setlength{\unitlength}{1\textwidth}
\begin{picture}(1,0.36)
  \put(0.005,-0.01){\includegraphics[width=0.99\textwidth]{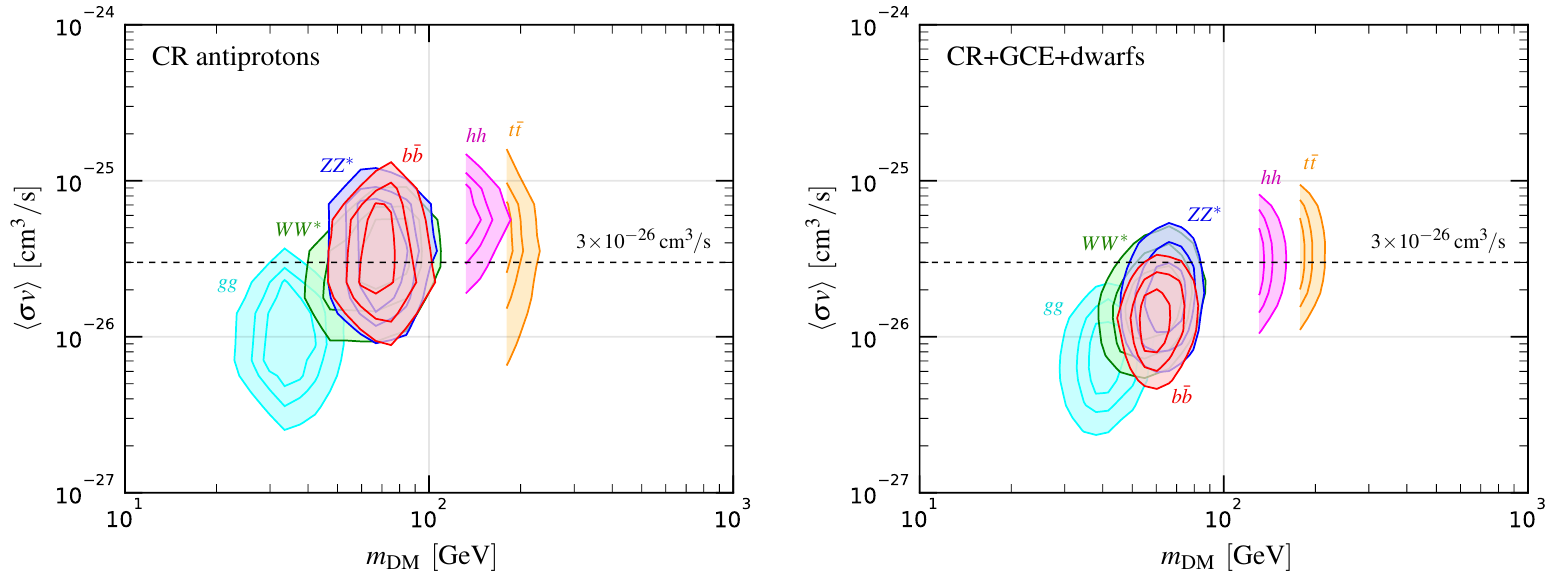}}
\end{picture}
\caption{Left: Best-fit regions (1--3$\sigma$ frequentist contours) for a dark-matter component of the 
cosmic-ray antiproton flux assuming 100\% annihilation into $gg$~(cyan), $WW^{(*)}$~(green), $b\bar b$~(red), 
$ZZ^{(*)}$~(blue), $hh$~(pink) and $t\bar t$~(orange).
Right:~Same for a combined fit of cosmic-ray antiprotons, the gamma-ray Galactic center excess and gamma-ray observations of dwarf spheroidal galaxies.
The figure is taken from Ref.~{\protect\refcite{Cuoco:2017rxb,Cuoco:2017okh}}.
}
\label{fig:fitreg}
\end{figure}
%                                      \         |
%                                        \       |
%                                          \     |
%=====================

Intriguingly, dark matter with very similar properties\footnote{See \eg~Ref.~\refcite{Calore:2014nla} showing a pattern very similar to the best-fit regions in the left panel of figure~\ref{fig:fitreg}.} has also been found to fit the gamma-ray Galactic center excess seen in the Fermi-LAT data~\cite{Goodenough:2009gk,Vitale:2009hr,Hooper:2010mq,Hooper:2011ti,Abazajian:2012pn,Hooper:2013rwa,Gordon:2013vta,Abazajian:2014fta,Daylan:2014rsa,Calore:2014xka,TheFermi-LAT:2015kwa,Karwin:2016tsw}.${}^,$\footnote{The origin of the excess is, however, controversially discussed in the literature~\cite{Petrovic:2014uda,Petrovic:2014xra,Cholis:2015dea,Bartels:2015aea,Lee:2015fea,Fermi-LAT:2017yoi,Leane:2019xiy,Chang:2019ars,Leane:2020pfc}.}
In fact, a simultaneous fit of antiprotons and gamma-rays reveals very good compatibility of the two observations when interpreted as a signal from dark matter~\cite{Cuoco:2017rxb}. This is, in particular, true for annihilation into a pair of $b$-quarks, $W$- and $Z$-bosons, as well as Higgses. Annihilation into gluons is slightly disfavored as both signals individually prefer somewhat different regions in the dark-matter mass while annihilation into top-quarks fits neither observation to the same level as the lighter final states. 
The combined fit (including a further likelihood contribution from gamma-ray observations of dwarf spheroidal galaxies~\cite{Fermi-LAT:2016uux}) is shown in the left panel of figure~\ref{fig:fitreg}. 
The preference for somewhat smaller annihilation cross sections in the combined fit is driven by the gamma-ray observations. It takes advantage of the relatively large uncertainties in the local dark matter density (taken into account in the fit) towards larger values.

While figure~\ref{fig:fitreg} shows the best-fit regions for individual final state channels, 
realistic dark-matter models often provide annihilation in an admixture of final states.
In Ref.~\refcite{Cuoco:2017rxb} we have performed a dedicated analysis of 
the Higgs portal dark-matter model~\cite{Silveira:1985rk}, where (for dark-matter masses below the Higgs mass)
annihilation proceeds via an intermediate Higgs in the $s$-channel. In this case, the composition of final state particles
corresponds to their coupling to the Higgs and the kinematically accessible phase space. Hence, it is solely a function of the dark-matter mass. 
The analysis shows that the model provides an excellent fit to the antiproton excess (alone and in combination with the gamma-ray Galactic center excess; see also Ref.~\refcite{Cuoco:2016jqt}) 
for a dark-matter mass around 60\,GeV, where annihilation into $b\bar b$ and $WW^*$ dominate. 
This is interesting, since independent of this observation, the resonantly enhanced region, $m_\text{DM}\simeq m_h/2$, is one of the few regions
the model survives a set of other model constraints~\cite{Cuoco:2017rxb}.

%=====================
%    \                                           |
%      \                                         |
%        \                                       |
\begin{figure}[t]
\centering
\setlength{\unitlength}{1\textwidth}
\begin{picture}(1,0.445)
      \put(0,-0.01){\includegraphics[width=1\textwidth]{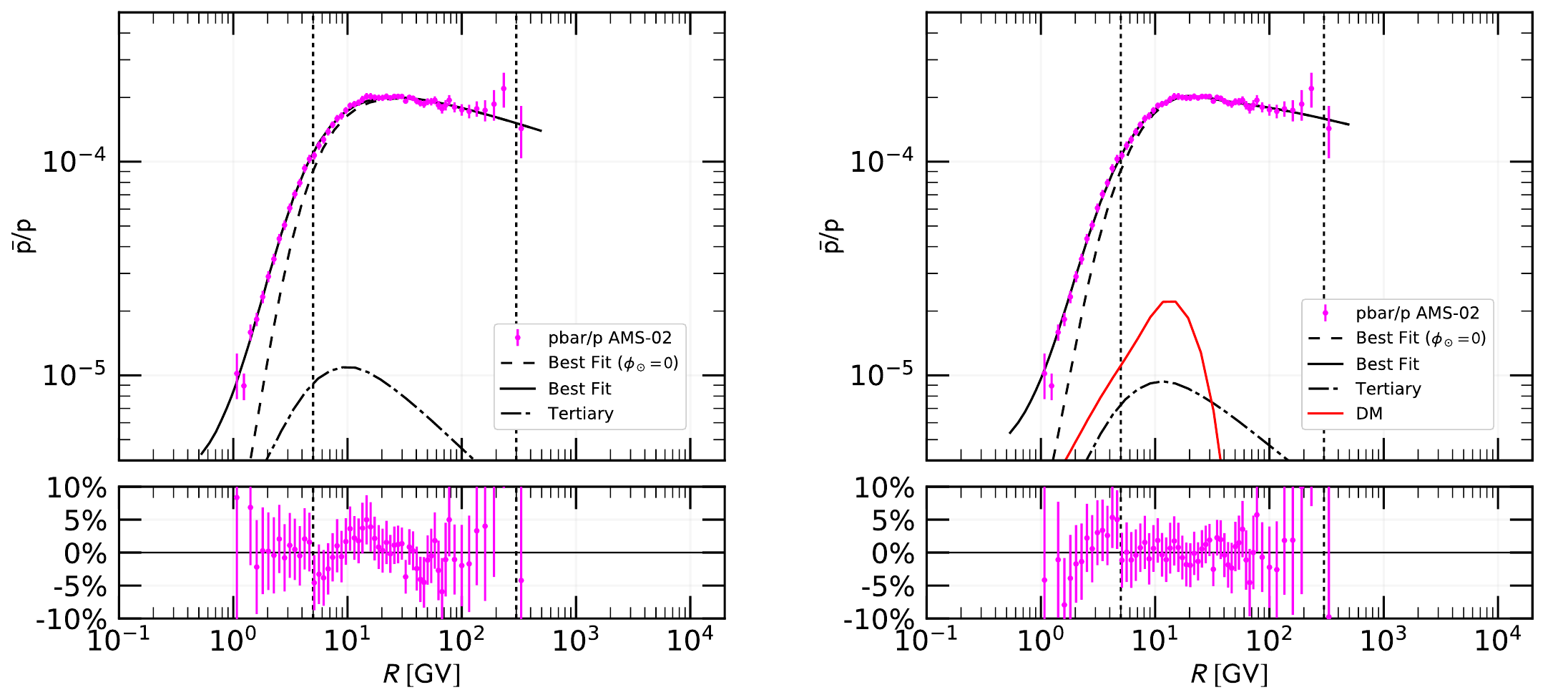}}     
    \end{picture}
	\caption{Antiproton-over-proton ratio for the respective best-fit points without (left) and with (right) a dark-matter component (annihilation into $b\bar b$).
	The data is fitted in the range $R=(5\!-\!300)$\,GV (between the dotted lines).  
	The figure is taken from Ref.~{\protect\refcite{Cuoco:2019kuu}}.
	}
  \label{fig:pbar_spectra_fits}
\end{figure}
%                                      \         |
%                                        \       |
%                                          \     |
%=====================

%===================================================================
\section{Systematic uncertainties of the antiproton excess}\label{sec:uncer}
%===================================================================

In this section, we take a closer look at the origin of the discrepancy in the signifi\-cance of the 
antiproton excess mentioned in the last section. This motivates us to take a deeper look into 
systematic uncertainties. 
As shown in figure~\ref{fig:pbar_spectra_fits} (taken from Ref.~\refcite{Cuoco:2019kuu})
the preference for dark matter arises from a relatively subtle effect,
namely a spectral feature around 10--20\,GV, best seen in the residuals of the left panel.
As shown in the right panel, the tension with data can be reconciled (\cf~the residuals) 
by a spiky contribution from dark matter that amounts to around 10\% of the total flux only
(red solid curve in the main plot on the right). Note that the relative uncertainty on the measurement
of the $\bar p / p$ flux ratio is around 4\%.
However, the global fit analysis induces a complex network of constraints on the free parameters and, 
hence, requires a careful assessment of the various sources of 
systematics and their possible correlations.

In the following, we discuss systematic errors that could have `faked' the signal.
Notably, this concerns uncertainties in the production cross sections of secondary antiprotons as well as 
correlations in the AMS-02 systematic error.

%-----------------------------------------------------------------------------------------------------------
\subsection{Uncertainties in the antiproton production cross sections}\label{sec:xs}
%-----------------------------------------------------------------------------------------------------------

The inclusive antiproton production cross sections entering
the secondary antiproton source term, eq.~\eqref{eq:sec_sourceTerm}, cannot be computed from first principles.
There are, however, two different frameworks to utilize measurements of the 
antiproton production cross sections at laboratory experiments for making corresponding predictions. 

The historically first one is the introduction
of an analytic parametrization of the fully differential Lorentz invariant cross section
using a functional form that is theoretically motivated by the scaling hypothesis~\cite{Taylor:1975tm}.
Its parameters are then fitted to data. This framework has first been employed in Ref.~\refcite{Tan:1983de}.
While, at the time, its use involved a significant degree of extrapolation into unconstraint territory, in particular towards high energies, 
the framework underwent a continuous refinement supported by new data that became available~\cite{Duperray:2003bd,diMauro:2014zea,Kappl:2014hha,Winkler:2017xor,Korsmeier:2018gcy}.
Notably, this involved a careful assessment of uncertainties~\cite{diMauro:2014zea}, a dedicated modeling of the hyperon and antineutron contributions~\cite{Kappl:2014hha}
and the introduction of scaling violations towards high energies~\cite{Winkler:2017xor}. Most recently, the parametrizations of Refs.~\refcite{diMauro:2014zea} and~\refcite{Winkler:2017xor}
have been re-fitted including new data from the NA61 and LHCb experiments in Ref.~\refcite{Korsmeier:2018gcy}.

The second path to computing the antiproton source term is the use of Monte Carlo generators of hadronic interactions that have been calibrated on a wide range of data from accelerator experiments.
This approach has, for example, been pursued in~Refs.~\refcite{Simon_Antiproton_CS_Scaling_1998,Donato:2001ms} utilizing DTUNUC and more recently in Ref.~\refcite{Kachelriess:2015wpa} using QGSJET-II-04 and  EPOS-LHC which underwent a re-tuning based on LHC data~\cite{Pierog:2013ria}. While this approach is favorable for high energies, 
it is rather unwarranted for kinetic energies relevant here, $T_{\bar p}\lesssim 10$\,GeV, as the underlying theoretical framework neglects several reaction mechanisms of potential importance in this regime, like~\eg~Reggeon exchanges or intranuclear cascading~\cite{Kachelriess:2015wpa}.

A comparison of the resulting source terms from the process $p p \to \bar p X$ for different parametrizations is shown in the left panel of figure~\ref{fig:pbar_sec_source}.
It includes the parametrizations of Refs.~\refcite{diMauro:2014zea} and~\refcite{Winkler:2017xor}
after (before) the parameter reevaluation in Ref.~\refcite{Korsmeier:2018gcy} [denoted by `Param.~I'  (`di Mauri') and `Param.~II' (`Winkler'), respectively] as well as the model of Ref.~\refcite{Kachelriess:2015wpa}
(denoted by `KMO'). Furthermore, it shows the $2\sigma$ uncertainties derived in Ref.~\refcite{Korsmeier:2018gcy}. 
The reevaluated models are roughly consistent with each other within errors. While, towards high energies, the parametrization of Refs.~\refcite{diMauro:2014zea} changed significantly with the parameter update, the one of Ref.~\refcite{Kachelriess:2015wpa} was only affected mildly. Notably, it is consistent with Monte Carlo based prediction of Ref.~\refcite{Kachelriess:2015wpa} at high energies. As expected, large discrepancies emerge between the latter and all analytic parametrizations at low energies.

%=====================
%    \                                           |
%      \                                         |
%        \                                       |
\begin{figure}[t]
\centering
\setlength{\unitlength}{1\textwidth}
\begin{picture}(0.96,0.43)
      \put(0,-0.04){\includegraphics[width=0.5\textwidth]{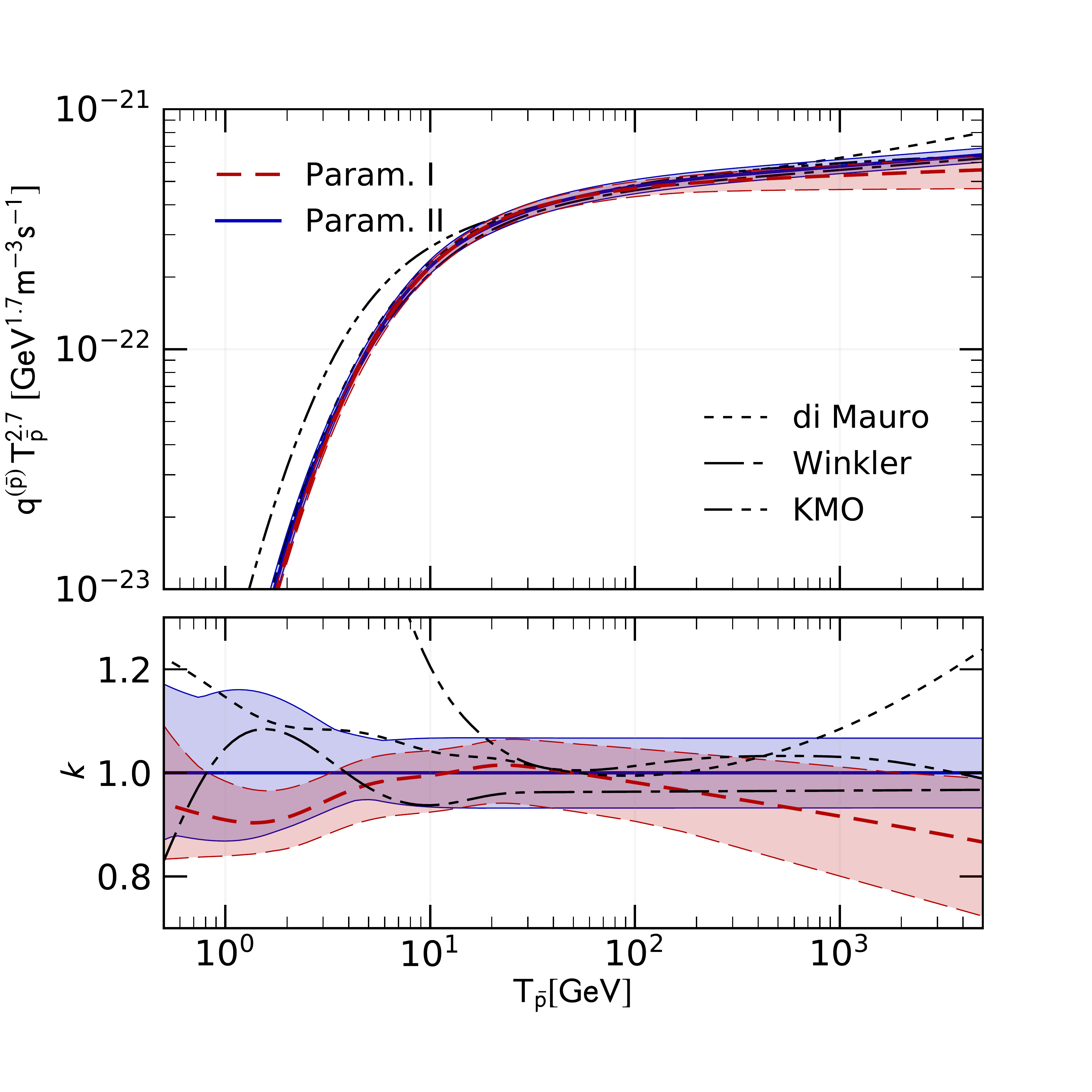}}     
      \put(0.505,-0.04){\includegraphics[width=0.5\textwidth]{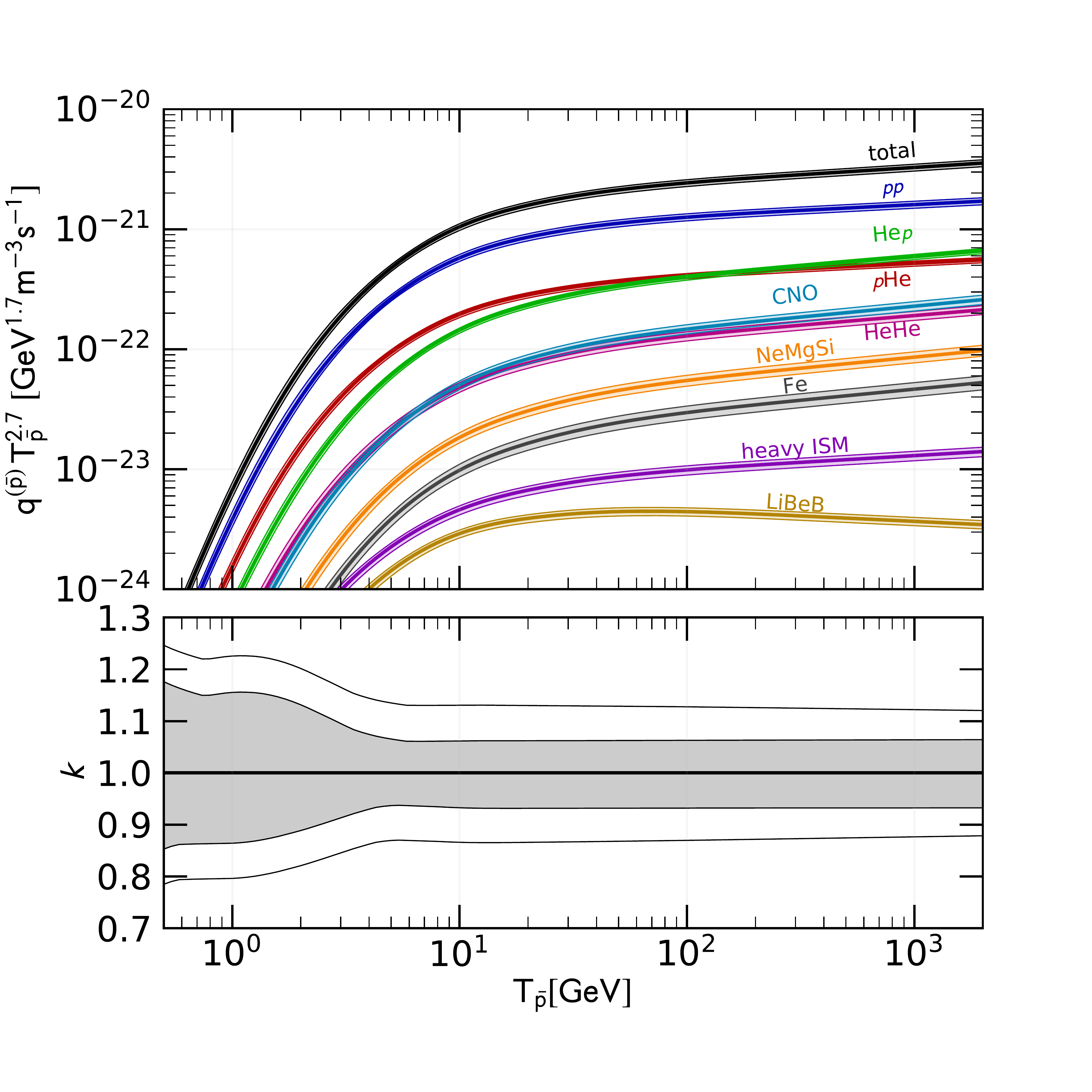}}     
    \end{picture}
	\caption{Secondary antiproton source term as a function of the antiproton kinetic energy.
	Left panel: Contribution from $pp\to\bar p +X$ for the parametrizations of Refs.~{\protect\refcite{diMauro:2014zea}} and~{\protect\refcite{Winkler:2017xor}}
	with re-fitted parameters~\cite{Korsmeier:2018gcy} (denoted by `Param.~I' and `Param.~II', respectively) as well as with the original 
	parameters (denoted by `di Mauri' and `Winkler', respectively). Furthermore, the Monte Carlo model of Ref.~{\protect\refcite{Kachelriess:2015wpa}}
	is shown (denoted by `KMO'). Right panel: The total source term as well as all its sub-contributions.
  	The shaded bands in both panels report the $2 \sigma$ uncertainty for prompt $\bar p$ production. The additional outer lines in the 
  	bottom right panel (showing the relative uncertainty on the total source term) denote the uncertainty due to isospin effects and to hyperon decay.
	The figure is taken from Ref.~{\protect\refcite{Korsmeier:2018gcy}}.
	}
  \label{fig:pbar_sec_source}
\end{figure}
%                                      \         |
%                                        \       |
%                                          \     |
%=====================

The second most relevant antiproton production processes are $p$He and He\hspace{0.1ex}$p$ scattering. Note that in all parametrizations these processes have been modeled solely based on re-scaling and extrapolation from $pp$ scattering as well as proton scattering off heavier nuclei, such as C\@. Only recently, the LHCb experiment has provided measurements of the cross section for $p\,\text{He} \to \bar p +X$ utilizing the SMOG device~\cite{Aaij:2018svt}. As shown in Ref.~\refcite{Reinert:2017aga}, the measurements are in remarkable agreement with the predictions of the model from Ref.~\refcite{Winkler:2017xor}. This is an important result, providing confidence in the underlying framework. The data has first been included in a cross-section fit within the reevaluation of Ref.~\refcite{Korsmeier:2018gcy}.

The total secondary antiproton source term as well as all its sub-contributions are summarized in the right panel of figure~\ref{fig:pbar_sec_source} (for the parametrization Param.~I).
The corresponding relative uncertainty on the total source term is around 10--20\,\%. As the fit typically imposes strong correlations between the cross-section parameters~\cite{Winkler:2017xor,Korsmeier:2018gcy}, it is crucial to take into account the corresponding error correlation matrix in cosmic-ray fits. 

The effect of uncertainties in the antiproton production cross section on the tentative dark-matter signal has been studied using B/C to constrain diffusion~\cite{Reinert:2017aga,Cui:2018klo} 
as well as following the minimal network scenario described in section~\ref{sec:limits} (employing a joint fit of $p$, He and $\bar p / p$)~\cite{Cuoco:2019kuu}. The analysis in Ref.~\refcite{Reinert:2017aga} finds a significant reduction of the global significance of the excess to around $1\sigma$ once the above uncertainties are taken into account. In this study, correlations in the cross-section parameters have been translated into a covariance matrix for the rigidity bins of the measured flux in a Gaussian approximation. In Ref.~\refcite{Cuoco:2019kuu} we have followed the same approach as well as performed a combined fit of cross-section and propagation parameters. While both approaches give comparable results, the significance of the excess is much less affected by the inclusion of cross-section uncertainties in this setup. The different sensitivity to cross-section uncertainties in the two setups might in parts be explained by the total flux-normalization freedom in the setup of Ref.~\refcite{Cuoco:2019kuu} which effectively renders a fully correlated contribution to the cross-section uncertainty redundant.

%-----------------------------------------------------------------------------------------------------------
\subsection{Correlation in the AMS-02 data}\label{sec:corr}
%-----------------------------------------------------------------------------------------------------------

In the rigidity region of the antiproton excess (around $10\!-\!20$\,GV) the experimental
systematic uncertainties reported by AMS-02 dominate over the statistical ones. 
While these systematics are expected to be subject to sizeable correlations in rigidity,
so far this information has not been provided by the AMS-02 collaboration.
Hence, a common treatment has been to simply add statistical and systematic uncertainties in quadrature
considering them to be fully uncorrelated. However, such a treatment can effectively 
overestimate the uncertainties significantly, if a sizeable fraction of errors is correlated 
over a wide (or even the full) range of rigidities (the latter of which would amount to an overall normalization uncertainty).
In fact, the goodness-of-fit achieved in the analyses discussed above~\cite{Cuoco:2016eej,Cuoco:2017rxb,Cuoco:2019kuu}, 
typically $\chi^2/\text{d.o.f}\sim 0.2$, points in this direction. 

Furthermore, unaccounted error correlations on intermediate scales (\ie~over an intermediate number of rigidity-bins)
could induce unwanted features, falsely interpreted as an excess in the data.
Therefore a realistic assessment of the underlying correlations is of paramount importance
for data interpretation. 
A first attempt to model these correlations in a data-driven method has been provided in Ref.~\refcite{Cuoco:2019kuu} utilizing
covariance functions, characterized by a correlation length.
Notably, this analysis has revealed that correlations potentially have a dramatic effect on the significance of the antiproton excess.
The modeling in terms of covariance functions has been further refined in Refs.~\refcite{Derome:2019jfs,Boudaud:2019efq} by assigning
individual correlation lengths to each sub-contributions of the systematic error. 
However, the derivation of the individual correlation lengths had to rely on `educated guesses'~\cite{Derome:2019jfs}.

 %=====================
%    \                                           |
%      \                                         |
%        \                                       |
\begin{figure*}[t]
\centering
\setlength{\unitlength}{1\textwidth}
\begin{picture}(0.99,0.43)
 \put(-0.0055,-0.02){\includegraphics[width=0.53\textwidth, trim= {3.3cm 2.2cm 3cm 0.8cm}, clip]{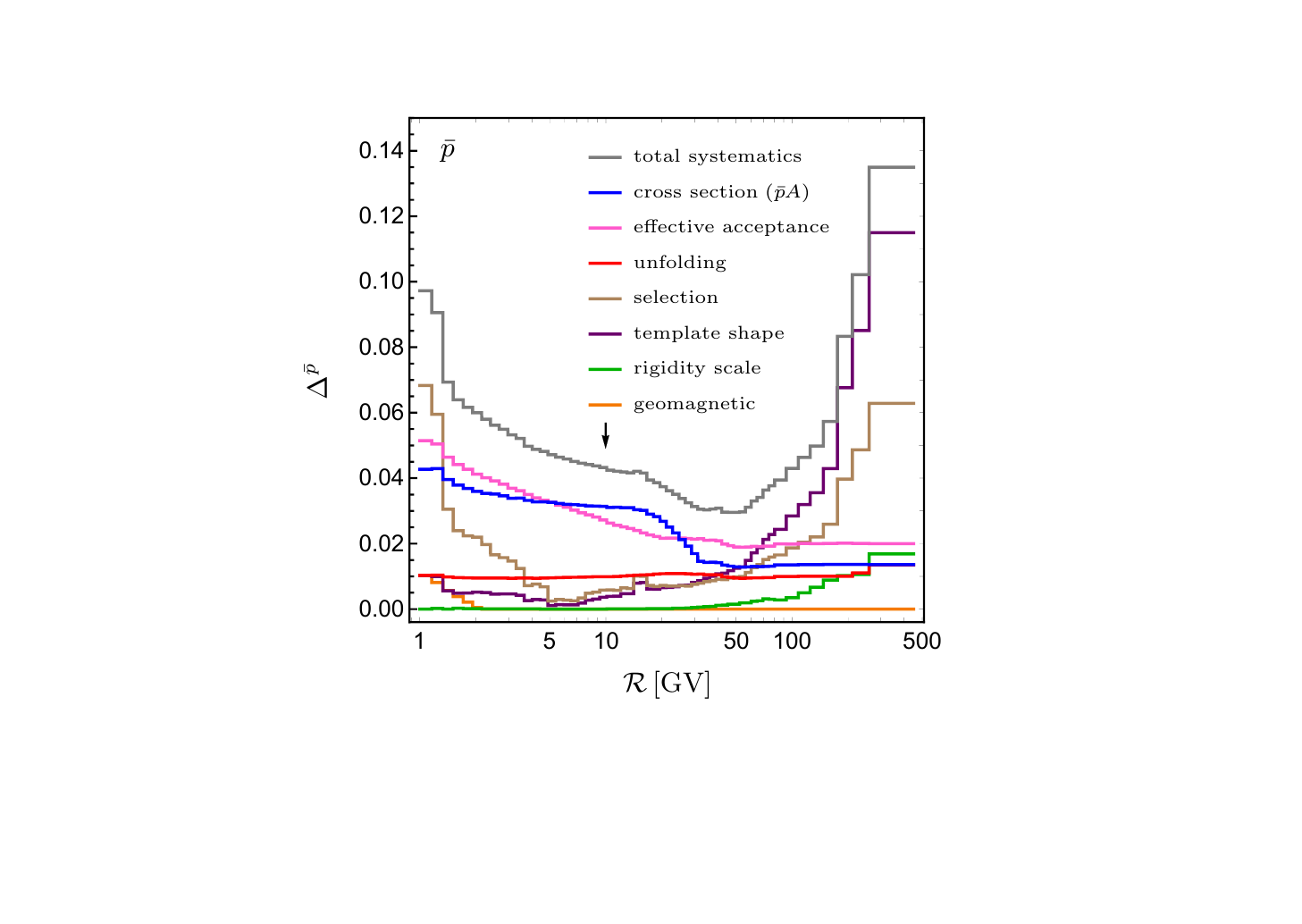}}
 \put(0.506,-0.02){\includegraphics[width=0.53\textwidth, trim= {3.3cm 2.2cm 3cm 0.8cm}, clip]{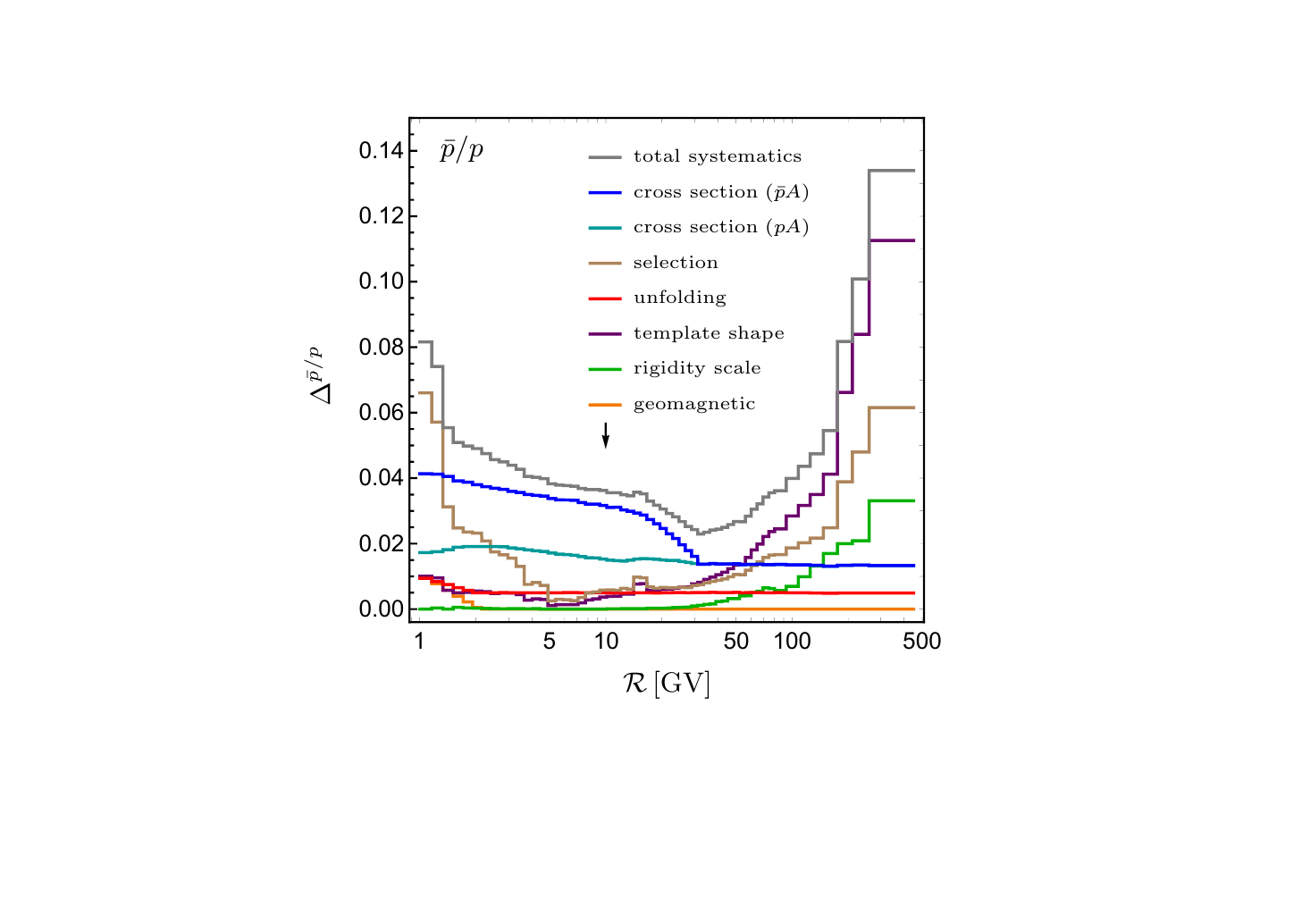}}
\end{picture}
\caption{
Reconstructed relative systematic uncertainties of the AMS-02 antiproton flux (left) and $\bar{p}/p$ flux ratio data (right). 
The individual contributions listed in the legend are ordered according to their size at 10\,GV, as indicated by the arrow.
The figure is taken from Ref.~{\protect\refcite{Heisig:2020nse}}.
\label{fig:systematicerrors}
}
\end{figure*}
%                                      \         |
%                                        \       |
%                                          \     |
%=====================

While this is \emph{a priori} due to the limited public availability of information from the collaboration, some
correlations entering the analysis have, in fact, not been rigorously computed until very recently~\cite{Heisig:2020nse}.
This concerns the uncertainties in the cross sections for (anti)proton absorption in the detector material 
the measured fluxes are corrected for. As figure~\ref{fig:systematicerrors} shows, these turn out to be the dominant
uncertainties around 10--20\,GV, \ie~in the region most relevant for the excess, both for the antiproton flux and the $\bar{p}/p$ flux ratio.
At the same time, measurements of the involved nucleon-carbon\footnote{The AMS-02 detector is dominantly composed of carbon.} absorption cross sections
from laboratory experiments often date back to the 1950s to 1980s and may involve  
unaccounted systematics.
In Ref.~\refcite{Heisig:2020nse} we, hence, performed a detailed reevaluation of these cross sections
within the Glauber-Gribov theory of inelastic scattering~\cite{Glauber1959,Gribov:1968jf,Pumplin:1968bi}. The theory links the nuclear absorption cross section 
to the nucleon-nucleon scattering cross sections and nuclear density functions, that are subject to independent experimental measurements.
It thus enables a welcomed redundancy in the parameter determination that we have exploited in a global fit of the data, see figure~\ref{fig:pbCxs}.
Most significantly, the fit allowed us to compute the correlations in the absorption cross-section uncertainties. Note that the respective correlation matrix cannot be characterized by a constant correlation length. 
Compared to the estimate in Ref.~\refcite{Boudaud:2019efq}, it tends to provide stronger large-scale correlations, \ie~correlations over a wider range of rigidity bins. 

%=====================
%    \                                           |
%      \                                         |
%        \                                       |
\begin{figure*}[t]
\centering
\setlength{\unitlength}{1\textwidth}
\begin{picture}(0.52,0.35)
 \put(-0.0055,-0.03){\includegraphics[width=0.55\textwidth, trim= {3.3cm 2.2cm 3cm 2cm}, clip]{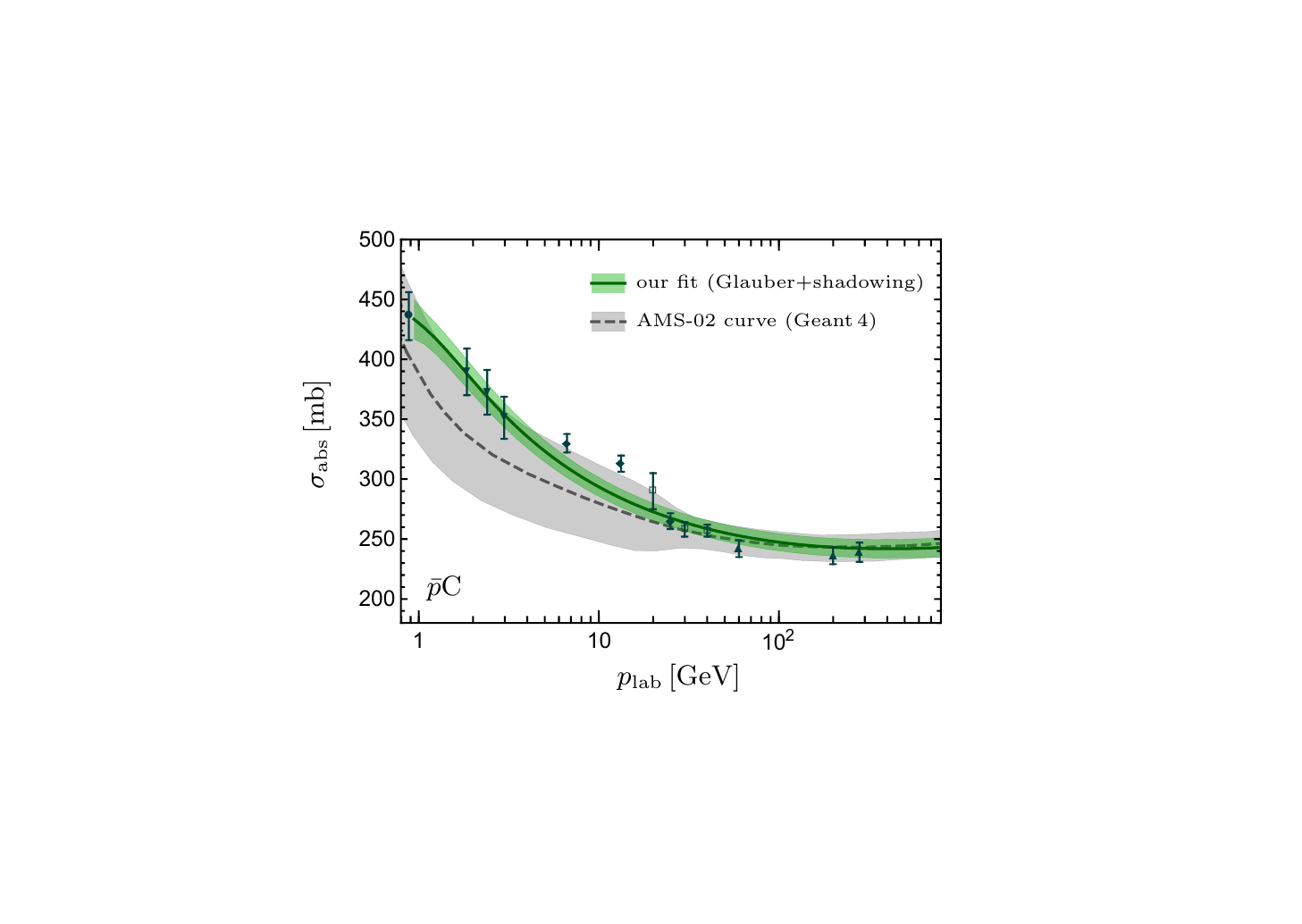}}
\end{picture}
\caption{
Absorption cross section for $\bar p$C as a function of the antiproton momentum $p_\text{lab}$.
The solid dark green curve and green shaded band denote the best-fit cross section and its $1\sigma$ uncertainty, respectively, within 
our global fit. Data points (containing $1\sigma$ error bars) of different experiments are denoted by individual symbols.
For comparison, the corresponding cross section used in the AMS-02 analyses is shown (dashed curve and gray shaded band).
The figure is taken from Ref.~{\protect\refcite{Heisig:2020nse}}.
\label{fig:pbCxs}  
}
\end{figure*}
%                                      \         |
%                                        \       |
%                                          \     |
%=====================

For the antiproton flux, the second most relevant uncertainty stems from the effective acceptance, which derives from a comparison of the detector response between data and Monte Carlo simulation.
It is the only relevant uncertainty that exhibits strong correlations on short scales, \ie~between few neighboring bins~\cite{Boudaud:2019efq,Heisig:2020nse}.
This contribution is, however, absent in the $\bar p/p$ flux ratio as it cancels out.

Considering all mentioned uncertainties discussed so far allows us to revisit our conclusions
on the possible existence of a dark-matter contribution in cosmic-ray antiprotons. 
As shown in Ref.~\refcite{Heisig:2020nse}, when fitting the $\bar p$ flux, the preference for a signal is entirely gone
once uncertainties from the antiproton production cross section and the correlations in the AMS-02
data are taken into account. This is independent of whether the propagation model is constrained
in a joint fit together with proton and He (minimal network setup of Ref.~\refcite{Cuoco:2019kuu}) or through a fit of the B/C flux ratio (setup of Ref.~\refcite{Reinert:2017aga}).
In particular, the effective acceptance error, which exclusively introduces sizable small-scale correlations, appears to be capable of `absorbing' the sharp spectral feature seen as the excess.
However, when fitting $\bar p/p$, additional freedom in the diffusion model at low rigidities is needed
[in this case parametrized by $\eta\neq1$; \cf~eq.~\eqref{eq:diff}] to eliminate the preference for dark matter. 
This appears to be in line with the conclusions drawn in~Ref.~\refcite{Boudaud:2019efq}, showing that
antiprotons can be well fitted when allowing either for $\eta\neq1$ or a low-rigidity break in diffusion 
(while taking into account the above discussed uncertainties, although estimated slightly differently).
However, we emphasize that both analyses do not exclude the possibility of a dark-matter signal. 
Note that the correlations in the effective acceptance error (derived in Ref.~\refcite{Heisig:2020nse} in a data-driven approach) plays a crucial role in the
analysis entering all measured (absolute) fluxes. First-hand information about this contribution would hence be a valuable input to the discussion.

%===================================================================
\section{Summary and conclusions}\label{sec:sum}
%===================================================================

Cosmic-ray antiprotons constitute a remarkable diagnostic tool for the study of astroparticle physics' processes in our Galaxy.
Certainly, the bulk of the measured antiprotons are consistent with a secondary origin
arising from collisions of primary cosmic-ray nuclei with the interstellar gas.
However, with new data from the AMS-02 experiment, uncertainties are -- for the first time -- at the percent level,
equipping us with encouraging prospects to pinpoint a possible primary component
of antiprotons, either of astrophysical origin or of exotic nature, such as dark-matter annihilation.

In this article, we reviewed recent developments in the search for the latter. 
For heavy dark matter, $m_\text{DM}>200\,\text{GeV}$,
a joint fit of propagation and dark-matter parameters in the `minimal network' scenario (using $\bar p /p $, $p$ and He only)
has led to strong limits on the annihilation cross section excluding the canonical value, $3\times 10^{-26}\,\text{cm}^3/\text{s}$,
up to dark-matter masses around 800\,GeV for a variety of non-leptonic channels.
This analysis considers uncertainties in the propagation model by profiling over its parameters in the fit.
Being largely insensitive to the choice of the dark-matter density profile in the Galaxy, the limits are robust and among the strongest 
current limits on self-annihilating dark matter. Similar results have been obtained using B/C to constrain propagation.

Probing smaller dark-matter masses, around or below 100\,GeV,
the data supports a possible hint for an annihilation signal.
This excess -- originating from a subtle spectral anomaly around $\mathcal{R}= 10\!-\!20\:\text{GV}$ --
has been seen by several groups using different analysis setups.
Interestingly, the signal is compatible with a thermal annihilation 
cross section for frozen-out dark matter as well as a dark-matter interpretation of the gamma-ray Galactic center excess. 
While the different studies agree on the preferred dark-matter properties, they draw very different conclusions
on the significance of the excess, calling out for further scrutiny of the finding.

Several systematic uncertainties have been assessed that could have `faked' the signal. 
An important one comes from our limited knowledge of the secondary antiproton production cross sections 
in the kinematic regime relevant for the antiproton source term. With new experimental results (most recently, LHCb data on $p \,\text{He}\to \bar p +X$)
and recent progress on the modeling of the hyperon and antineutron contributions as well as scaling violations,
their description has significantly improved over the last couple of years. 
However, the models largely rely on input from old data, in particular in the low-energy regime. The involved experimental systematics, \eg~regarding the hyperon contribution,
are often poorly known. 

Another important aspect is the presence of correlations in the experimental errors of the AMS-02 measurements, which have not been provided by the collaboration. 
In the rigidity region of interest, around 10--20\,GV, errors of the antiproton flux (or flux-ratio) are dominated by systematics. 
The most relevant ones are uncertainties in the cross sections for cosmic-ray absorption in the AMS-02 detector
the measured fluxes are corrected for. A careful reevaluation of these absorption cross sections, that led to the computation of the corresponding error
correlations, has only been performed and made publicly available recently. 

Remarkably, the consideration of all these uncertainties eliminates the statistical preference for an additional contribution from dark matter in various analysis setups.
While these findings cast severe doubts on the robustness of the excess, the situation is not fully conclusive. 
Correlated uncertainties in the `effective acceptance' (that have as well been incorporated but could only be estimated on the basis of the limited information publicly available)
and the modeling of the diffusion coefficient at low energies also play an important role. They motivate further investigations.

To fully exploit the wealth of data from AMS-02 requires various improvements both on the experimental and theoretical sides. 
First, the provision of the covariances for key systematic errors such as the effective acceptance would settle doubts in their estimates done outside the collaboration.
Secondly, taking advantage of the recent progress in the computation of cosmic-ray absorption cross sections, a reevaluation of the measured fluxes and their uncertainties could be performed by the collaboration. Thirdly, to further gain sensitivity to the low-rigidity behavior of the antiproton spectrum, solar modulation may have to be incorporated beyond an improved force-field approximation
making use of the time-resolved data provided. 

Finally, an independent test of the excess and its dark-matter interpretation can only be done by a multi-messenger approach. 
In particular, observations of low-energy antideuterons provide a low-background search channel with promising prospects for future experiments~\cite{Aramaki:2015pii,Korsmeier:2017xzj,vonDoetinchem:2020vbj}.

%===================================================================
\section*{Acknowledgments}
%===================================================================

I thank Alessandro Cuoco and Martin W.~Winkler for valuable comments on the manuscript.
I acknowledge support from the F.R.S.-FNRS, of which I am a postdoctoral researcher.

\end{document}